# XPS and DFT study of Sn incorporation into ZnO and TiO$_2$ host matrices by pulsed ion implantation

**D.A. Zatsepin[1,2], D.W. Boukhvalov[3,4,*], E.Z. Kurmaev[1,2], I.S. Zhidkov[2], S.S. Kim[5], L. Cui[6], N.V. Gavrilov[7], S.O. Cholakh[2]**

[1] M.N. Miheev Institute of Metal Physics, Ural Branch of Russian Academy of Sciences, 18 Kovalevskoj Str., 620990 Yekaterinburg, Russia
[2] Technological Institute, Ural Federal University, Mira Street 19, 620002 Yekaterinburg, Russia
[3] Department of Chemistry, Hanyang University, 17 Haengdang-dong, Seongdong-gu, Seoul 133-791, Korea
[4] Theoretical Physics and Applied Mathematics Department, Ural Federal University, Mira Street 19, 620002 Yekaterinburg, Russia
[5] School of Materials Science and Engineering, Inha University, Incheon 402-751, Republic of Korea
[6] Shenzhen Graduate School, Harbin Institute of Technology, Shenzhen 518055, P.R. China
[7] Institute of Electrophysics, Russian Academy of Sciences–Ural Division, 620016 Yekaterinburg, Russia

*Bulk and thin films ZnO and TiO$_2$ samples were doped with Sn by pulsed ion implantation and studied by means of X-ray photoelectron core-level and valence band spectroscopy as well as density functional theory calculations for comprehensive study of the incorporation of Sn. XPS spectral analysis showed that isovalent Sn cation substitution occurs in both zinc oxide ($Sn^{2+} \rightarrow Zn^{2+}$) and titanium dioxide ($Sn^{4+} \rightarrow Ti^{4+}$) for bulk and film morphologies. For TiO$_2$ films, the implantation also led to occupation of interstitials by doped ions, which induced the clustering of substituted and embedded Sn atoms; this did not occur in ZnO:Sn film samples. Density functional theory (DFT) formation energies were calculated of various incorporation processes, explaining the prevalence of substitutional defects in both matrices. Possible mechanisms and reasons for the observed trends in Sn incorporation into the ZnO and TiO$_2$ matrices are discussed.*

## 1. Introduction

During recent decades metal-doped zinc and titanium oxides have been studied intensively (see e.g. Refs. [1-7]) because of their great importance in technological applications. The reason for that derives from understanding that the bundle of physical and chemical properties of these doped oxides are determined by applied fabrication method and, hence, the final microstructure of a material. When doping an oxide, a dopant may assume different crystallographic positions inside the host matrix, depending on whether isovalent, isostructural or interstitial doping occurs under concrete material processing [8], and the oxygen partial density-of-states contribution to the electronic properties will differ accordingly. However, the final microstructure of doped oxide is not determined only by the sintering technique used; the role of the host matrix type is also important [9], causing as a reply to technological treatment, an appearance of microheterogeneity and imperfection areas, unstable phase separation of the final material, host matrix structural defects, etc. All these peculiarities strongly impede the accumulation and analysis of objective research data and have hindered the development of novel functional materials based on the zinc and titanium oxide systems.

To controllably modify most of ZnO's properties, the doping by Ga, Ge, Al, Pb and Sn is usually employed [10]. Among these, the group IV dopants (Pb and Sn) are typically considered the most important because they can potentially provide a high concentration of free charge carriers, strongly affecting ZnO's conductivity and modifying the overall band structure; these dopants can become incorporated into the ZnO host as singly or doubly ionized donors. The process of doping with Sn is not completely understood; there are two opposite points of view regarding the incorporation of Sn into ZnO. One school of thought is that $Sn^{4+}$ substitutes $Zn^{2+}$, leading to the formation of $SnO_2$ clusters and Zn vacancies; this theory accounts well for the differences in field emission characteristics between ZnO:Sn and undoped ZnO [11]. As an alternative, $Sn^{2+} \rightarrow Zn^{2+}$ isovalent substitution has also been reported and proved with the help of X-ray diffraction and X-ray photoelectron spectroscopy techniques (see e.g. Refs. [12-15]). Neither of these competing models invokes reconstructive conversion of the host matrix structure after doping.

For Sn doping of $TiO_2$ the situation is different. In the most cases an isovalent $Sn^{4+} \rightarrow Ti^{4+}$ substitution has been reported, and has been mentioned as the only one that is energetically favourable [6-7, 16]. However, in contrast to ZnO, some researchers have indicated that there is a total reconstruction of the $TiO_2$ structure after doping [5]; unfortunately, they did not prove whether this was a result of the sintering approach they applied, or of a subsequent thermal treatment that they applied to the $TiO_2$:Sn. Also, there has been no complete analysis of both oxide matrices regarding how the morphology of the sample – film or bulk – contributes to the incorporation mechanism. So the questions mentioned above need further clarification.

Herein we report on an X-ray photoelectron spectroscopy study of ZnO and $TiO_2$ host matrices doped with Sn ions by pulsed ion implantation, and related DFT calculations. For more comprehensive study, we performed measurements and calculations for both oxide hosts bulk ceramic (hereafter termed "bulk") and thin films.

**2. Sample Preparation, Experimental and Computational Details**

$TiO_2$ and ZnO bulk ceramic samples were made by hot pulsed pressing of appropriate ceramic powders obtained by electrical explosion of wires in oxygen-containing media. After processing in molds of 15 mm diameter at $7 \times 10^4$ N, the plates were sintered within 1 h at 1040 °C ($TiO_2$) and 1000 °C (ZnO). The samples were on average 13 mm in diameter and 1–2 mm in height; their density was 4.25 g cm$^{-3}$ for TiO$_2$ samples and 5.6 g cm$^{-3}$ for ZnO samples. We had verified phase composition by X-ray diffraction (XRD) technique, and the final TiO$_2$ host was found to be nearly all single-phase rutile (99.85%); the parameters of its tetragonal lattice were found to be a = 4.592 Å and c = 2.960 Å. Finally, the average crystallite size was determined to be >200 nm. The performed X-ray measurements for ZnO bulk ceramic samples demonstrate that the compacted material was single-phase zincite with hexagonal structure (a = 3.251 Å, c = 5.202 Å). The average determined ZnO crystallite size was about 200 nm.

TiO$_2$ coatings were prepared with the help of a sol–gel chemical technique in which titanium isopropoxide (97%), nitric acid (60%), and anhydrous ethanol had been used as the precursor, catalyst, and solvent, respectively. Purified and deionized water was used to hydrolyze the precursor mentioned, and all chemicals were used as received without any further purification. So the TiO$_2$ films were deposited on Si wafers (100) by means of a dip coating process, and 1-butanol was added to the coating sols in order to control their wettability and the viscosity. The substrates were ultrasonically cleaned within 30 min in acetone and ethanol in sequence. After that they were then washed with deionized water. The withdrawal rate of the substrate was nearly 4 mm s$^{-1}$. Finally, the as-prepared films were dried at room temperature and kept in an oven at 60 °C for 1 d to remove the remaining solvents completely; they were then annealed at 100 °C for 2 h. The anatase films were characterized by field emission scanning electron microscopy and atomic force microscopy to confirm that high-quality films were produced. Refer to Ref. 9 for full details on TiO$_2$ film synthesis and characterization. To deposit the ZnO thin films, a sapphire substrate (100) was ultrasonically cleaned in acetone and alcohol for 10 min, then rinsed in deionized water, and finally dried in N$_2$. The sapphire substrates were held at 250 °C for 90 min during the deposition, and the deposition was carried out at a working pressure of 2

Pa after pre-sputtering with Ar for 10 min. When the chamber pressure was stabilized, the radio frequency generator was set to 100 W. The growth rate of ZnO thin films was 3.4 nm min$^{-1}$ and the typical thin film thickness was 302 nm. The polycrystalline ZnO samples had a hexagonal structure with lattice parameters a = 3.250 Å and c = 5.207 Å.

Sn implantation of ZnO and TiO$_2$ thin films and bulk samples was made in a pulsed-repetitive mode at the residual gas pressure of $3 \times 10^{-3}$ Pa. The MEVVA-type ion source on a Sn-made cathode generated a beam with ion energy of 30 keV and pulse current density of 0.8 mA cm$^{-2}$; the pulse repetition rate of 12.5 Hz and pulse duration of 0.4 ms were used. An ion fluence (integrated flux over time) of $1 \times 10^{17}$ cm$^{-2}$ was achieved after 67 min of exposure. The average temperature of the samples did not exceed 300 °C during ion implantation.

X-ray photoelectron spectroscopy (XPS) measurements were made by using a PHI XPS Versaprobe 500 spectrometer (ULVAC–Physical Electronics, USA) with a quartz monochromator. XPS energy analyzer was supporting the working range of binding energies from 0 to 1500 eV with an energy resolution of $\Delta E \leq 0.5$ eV for Al *Kα* radiation (1486.6 eV). As in our previous studies, the samples were held in a vacuum (10$^{-7}$ Pa) for 24 h prior to measurement, and only samples whose surfaces were free from micro impurities (as determined by means of surface chemical state mapping attestation) were measured and reported herein. XPS spectra were recorded under monochromatized Al *Kα* X-ray emission with the spot size of 100 μm. An X-ray power load delivered to the sample was not more than 25 W in order to avoid x-ray stimulated decomposition of the sample. Under these conditions an XPS signal-to-noise ratios were achieved at least not worth than 10000:3. Primarily the spectra were processed using ULVAC-PHI MultiPak Software 9.3 and the residual background (BG) was removed by using the Tougaard approach with the Doniach–Sunjic line shape asymmetric admixture [17-19]. It is well known that most of the provided BG models are self-consistent and they use Doniach–Sunjic-type line shapes that are acceptable for most common XPS analyses. The advantage of retaining asymmetry in XPS data processing is usually strongly apparent when a Tougaard BG is used to remove the extrinsic contribution to the XPS spectrum of metal-like or metal-doped materials. Thus, in our case, it is a choice with a theoretical basis. After BG subtraction, the XPS spectra were calibrated using the reference energy of 285.0 eV for the carbon 1s core level. Using this exact sequence allows much better calibration due to the previous removal of outer contributions to the XPS line shape.

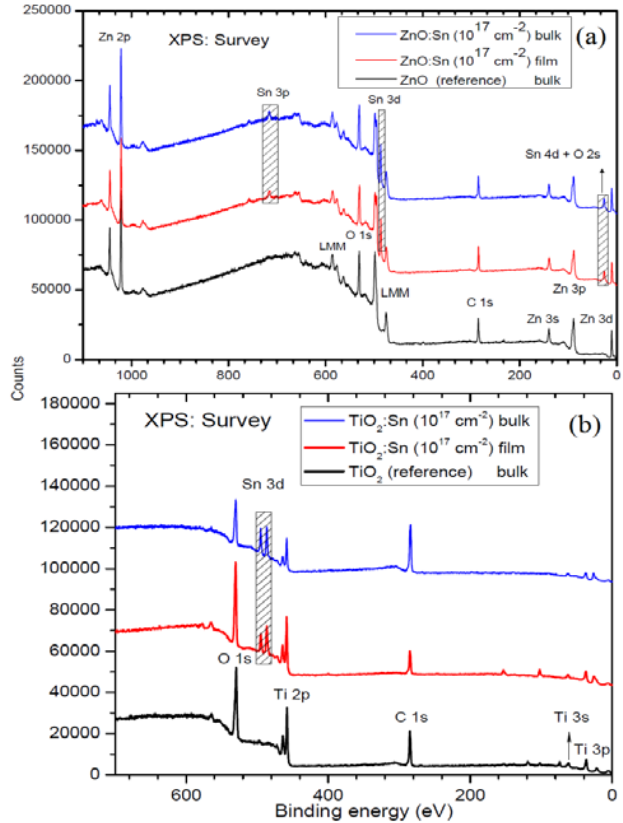

**Figure 1** XPS spectra for Sn-implanted ZnO (upper panel) and TiO$_2$ host matrices (lower panel) of bulk and film morphologies. Initial spectra of non-implanted ZnO and TiO$_2$ (bulk) are included for comparison.

Figure 1 displays the results of simple elemental analysis by means of XPS survey (wide-scan) spectra for the samples under study. It is clearly seen that both untreated and Sn-implanted ZnO and TiO$_2$ host matrices have no XPS signals from alien impurities, except for the Sn dopant that was injected by the ion beam. Carbon contamination was not exceeding ordinary carbon contamination for the films and bulk TiO$_2$ and ZnO samples that are used in photovoltaics (not more than 10 at. %) [20]. This analysis confirmed the declared empirical formulas and demonstrated the samples' high purity.

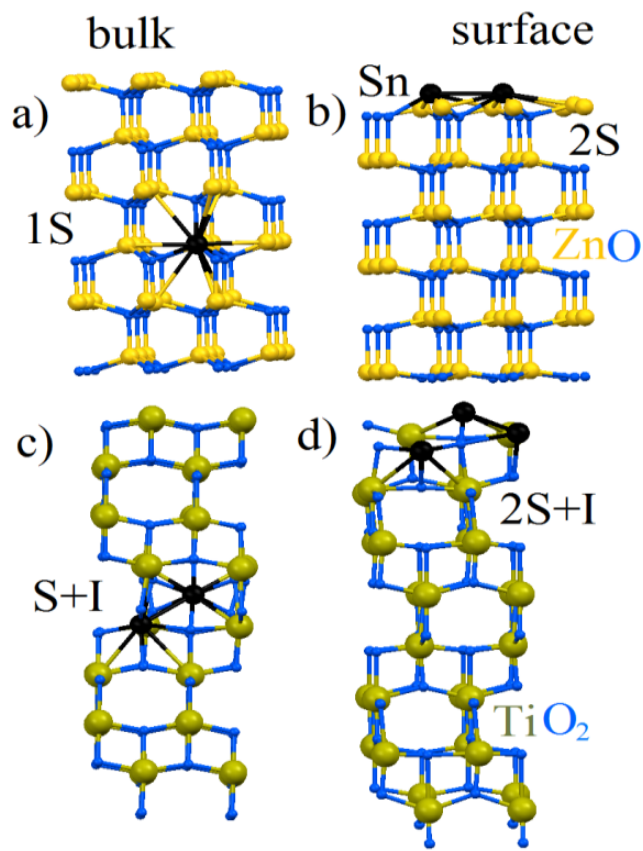

**Figure 2** Optimized atomic structure of studied defects in (a, b) ZnO and (c, d) TiO$_2$ hosts of (a, c) bulk and (b, d) surface.

Density functional theory (DFT) calculations were performed by using the SIESTA pseudopotential code [21,22], a technique that has been successful in related studies of impurities in semiconductors [23]. All calculations were made by using the Perdew–Burke–Ernzerhof variant of the generalized gradient approximation (GGA-PBE) [24] for the exchange-correlation potential. A full optimization of the atomic positions was carried out during which the electronic ground state was consistently found using norm-conserving pseudopotentials for the cores and a double-$\xi$ plus polarization basis of localized orbitals for Sn, Zn, Ti, and O. The forces and total energies were optimized with accuracies of 0.04 eV $\text{Å}^{-1}$ and 1.0 meV, respectively. For atomic structure calculations, we employed Sn and Ti pseudopotentials, treating the 3d electrons as localized core states. Calculations of formation energies ($E_{form}$) were performed by considering the supercell both with and without a given defect [23]. As a host for the studied defects in ZnO and $TiO_2$, the supercells consisting of 108 and 96 atoms respectively (Fig. 2) were used. Taking into account our previous modelling of transition metal impurities in semiconductors [16], we calculated various combinations of structural defects including single substitutional (1S; Fig. 2a) 3d impurities, pairs of substitutional impurities (2S; Fig. 2b) and their combinations with interstitial (I) impurities (S + I and 2S + I; Figs. 2c and 2d, respectively). Because in the case of thin films the contribution from the surface much more valuable we also examine surface effects we used slabs of ZnO and $TiO_2$ (Figs. 2b,d).

### 3. Results and Discussion

**3.1. XPS core-level and valence band spectra** To clarify the formal valence states of Sn dopants in ZnO and $TiO_2$ matrices after Sn implantation, XPS Sn 3d spectra were acquired from the samples under study (Fig. 3). From these spectra it could be seen clearly that the binding energy (BE) positions of the Sn $3d_{5/2}$ and Sn$3d_{3/2}$ bands of the ZnO:Sn sample were identical to those of the SnO reference (Fig. 3, upper panel). The Sn $3d_{3/2}$ signal was unusually intense due to the Auger Zn $L_3M_{45}M_{45}$ transition, which made it impossible to directly compare the spectral parameters of Sn$3d_{3/2}$ with those of the reference standard. Nevertheless, it was concluded that the formal valence state of Sn in ZnO:Sn sample is 2+ rather than 4+, and that it was isovalently incorporated into the ZnO host. The spectral dissimilarity between bulk and film samples in the BE region of 494–499 eV might be a consequence of the samples' different morphologies, which could lead to differences in the overlapping contribution of Auger Zn $L_3M_{45}M_{45}$ to the Sn $3d_{3/2}$ signal, because the $3d_{5/2}$ band was free from such overlapping and was identical in shape for both the bulk and film ZnO:Sn samples (Fig. 3). On the whole, our results do not contradict the previously reported results of Ref. [12].

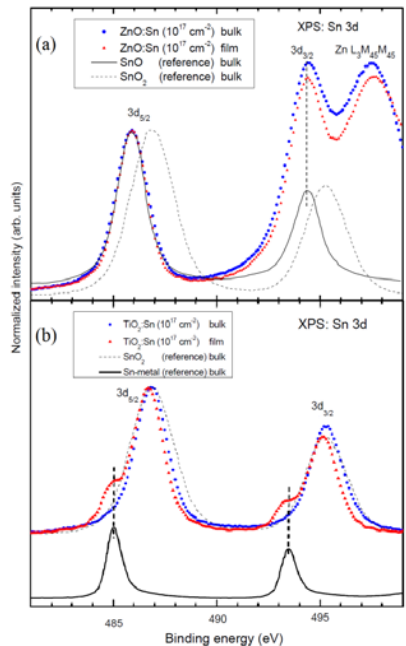

**Figure 3** XPS Sn 3d spectra for (upper panel) Sn-implanted ZnO and (lower panel) TiO$_2$ host matrices of bulk and film morphologies. Spectra of SnO (Sn$^{2+}$), SnO$_2$ (Sn$^{4+}$) and Sn metal (Sn$^0$) are included as external XPS standards.

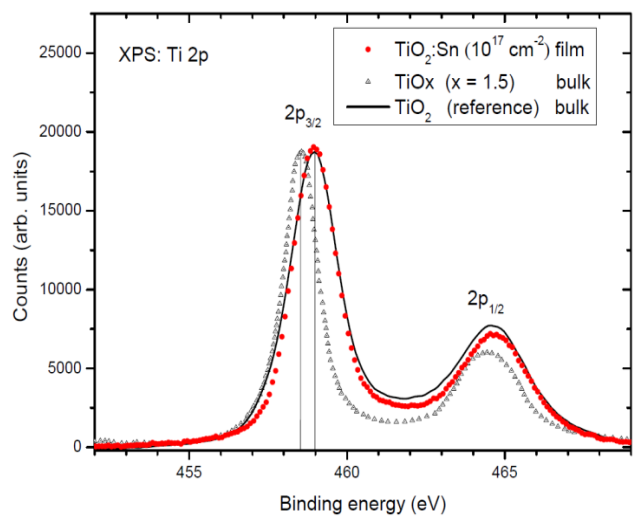

**Figure 4** XPS Ti 2p spectra for Sn-implanted $TiO_2$ and external XPS standards: reference $TiO_2$ and partially reduced $TiO_2$ — $TiO_x$ (x = 1.5).

The situation with $TiO_2$:Sn sample is dramatically dissimilar. Whereas the bulk sample exhibited an XPS Sn 3d spectral shape similar to that of the $SnO_2$ reference standard ($Sn^{4+}$), the spectrum of the film sample clearly had two additional subbands situated at 485 and 493.5 eV, respectively. According to the NIST XPS Database and Handbook of Photoelectron Spectroscopy, this feature might be the Sn $3d_{3/2}$ – Sn $3d_{5/2}$ signal of Sn metal; this is supported by the BE positions of the corresponding maxima in the XPS Sn 3d peak of pure Sn metal (Fig. 3, lower panel). Based on these findings, we supposed that in the bulk $TiO_2$:Sn sample, an isovalent incorporation of Sn occurred ($Sn^{4+} \rightarrow Ti^{4+}$), but in the film $TiO_2$:Sn not all injected Sn reacted with oxygen from the $TiO_2$ host sublattice, and thus some Sn accumulated as interstitials in a metal-like phase. In this situation, it cannot be ruled out that nonstoichiometric $SnO_x$ clusters might form, manifesting themselves as imperfections in the final $TiO_2$:Sn microstructure.

It is well known, that the $TiO_2$ films are potentially susceptible to ion beam reduction effects [25], so in order to reveal and control these effects in the future sample synthesis, the Ti 2p core-level spectra for Sn-implanted TiO$_2$ and references were recordered and analysed (see Fig. 4). According to [26], anatase and rutile polymorphs are identical with respect to their XPS Ti 2p core-level signal thus the Ti 2p spectrum shape transformation will be rather useful for the determination of Ti – O bonding imperfections but not to identify rutile or anatase phases. As one can see from this figure, there is no essential dissimilarities between XPS Ti 2p for reference TiO$_2$ and Sn-implanted TiO$_2$, whereas partially reduced TiO$_2$ (TiO$_x$) exhibits the core-level spectrum with a lower binding energies of Ti 2p$_{3/2}$ and Ti 2p$_{1/2}$ peaks as well as dissimilar spin-orbital splitting and intensity for Ti 2p$_{1/2}$ peak comparing with those for TiO$_2$ and TiO$_2$:Sn. From these findings we can assume that no reduction effects of TiO$_2$ host material occur during ion-beam synthesis as well as other side effects.

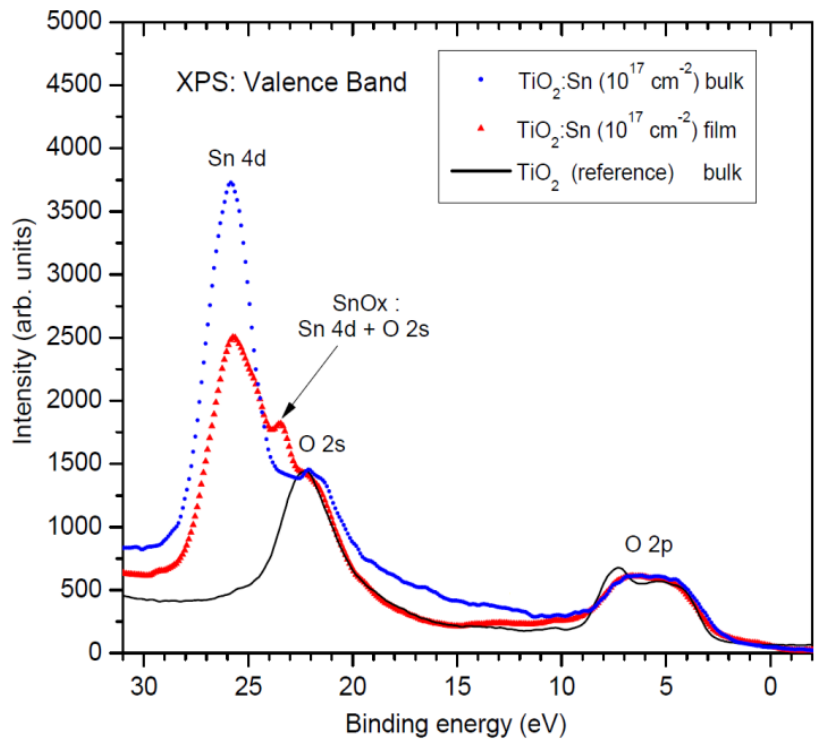

**Figure 5** XPS VB spectra for Sn-implanted TiO$_2$ host matrices of bulk and film morphologies. The appropriate TiO$_2$ (bulk) external XPS standard spectrum is added for comparison.

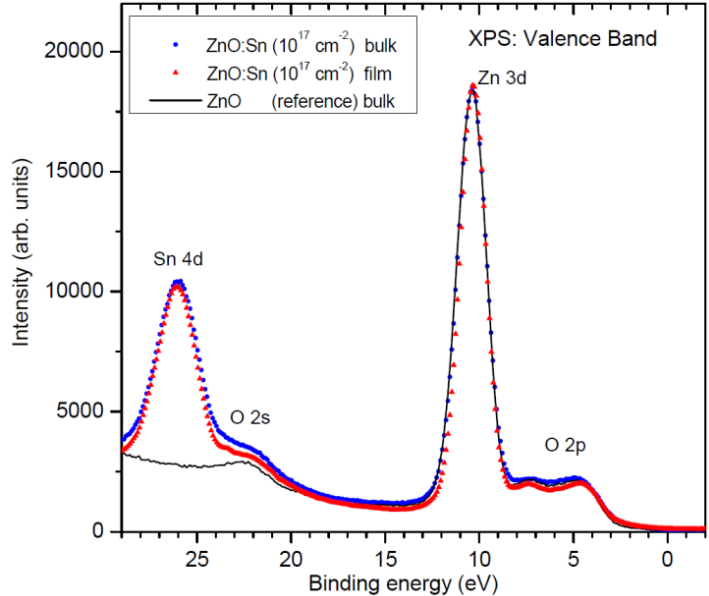

**Figure 6** XPS VB spectra for Sn-implanted ZnO host matrices of both bulk and film morphologies.

To clarify our supposition, XPS valence band (VB) spectra were acquired and analyzed (Fig. 5). The most dramatic difference between untreated and Sn-implanted $TiO_2$ was in the XPS VB BE region of 20–30 eV; here, the bulk $TiO_2$:Sn sample exhibited a strong Sn 4d signal and O 2s contributions. The most interesting feature was the appearance of an additional spectral band, located at approximately 23.7 eV, in the VB spectrum of the $TiO_2$:Sn film, which was absent in the spectra of both the $TiO_2$ reference standard and the bulk $TiO_2$:Sn. According to the NIST XPS Database this binding energy belongs to the hybridized Sn 4d – O 2s densities of states. De Padova et al. reported the same spectral feature, with BE of 24 eV, in their analysis of oxidized Sn foil [27]; they interpreted this XPS signal as a reply from $SnO_x$ clusters present in the structure of that material due to nonuniform oxidation. Taking into account our supposition, based on appropriate core-level spectra for $TiO_2$:Sn in bulk and film morphologies, and the data referenced just above, we can conclude that in the TiO$_2$:Sn film, the implantation caused Sn$^0$ species to become embedded in the host matrix and also introduced defects in the form of SnO$_x$. The latter was proved by our XPS valence band mapping analysis. In the XPS VB spectra of ZnO:Sn samples at 0-10 eV, no essential difference between Sn-doped bulk and film morphologies (Fig. 6) is found, evidencing that (isovalent) Sn incorporation occurs similarly in both.

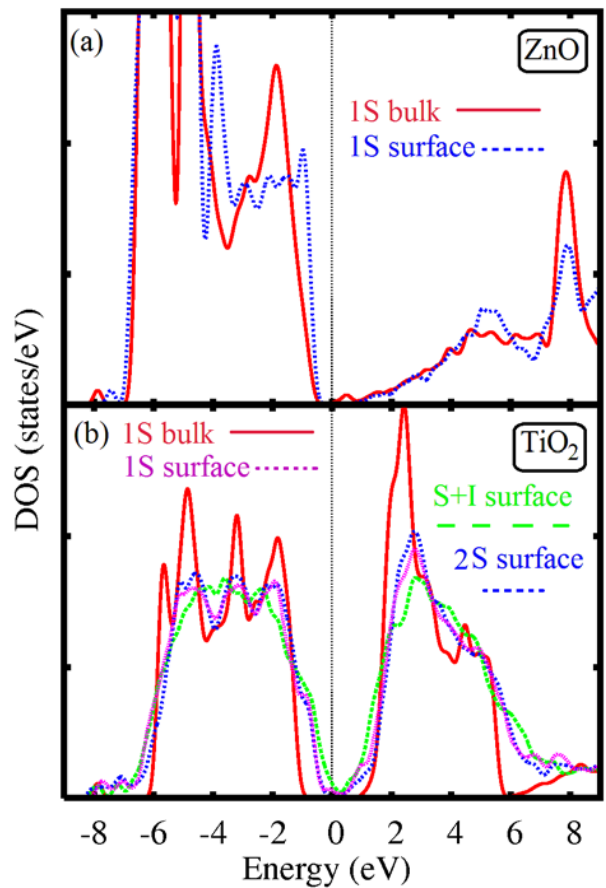

**Figure 7** Total densities of states for the most probable (see text) configurations of impurities in bulk and surface of ZnO (a) and TiO$_2$.

**3.2. DFT calculations** Table 1 lists the formation energies for bulk and surface defects, showing that Sn dopants become incorporated into the ZnO host only by means of substitution, and in the form of single impurity centres (1S). It is unlikely for paired Sn atoms (2S) to substitute the nearest-neighbourhood Zn atoms because of energy disadvantages. Similarly, it is also believed that most dopant atoms will reside on the sample surface. For titanium dioxide the situation is quite similar except that there is a possibility that paired Sn atoms substituting Ti atoms. An explanation for these DFT calculations regarding Sn dopant behaviour is based on the correlation between ionic radii for the fourfold zinc (formal valence state 2+) and sixfold titanium (formal valence state 4+); these ionic radii are 0.6 Å and 0.605 Å, respectively. As for Sn, the ionic radii for the corresponding nearest neighborhood pairs are 0.55 Å and 0.69 Å. This explains the relatively moderate values calculated for the formation energies, hence explaining the only substitution process in the bulk samples. For ZnO, the combination of substitutional and interstitial defects has similar formation energies for surface and bulk, in contrast to $TiO_2$, for which the surface formation energies of S+I and 2S+I clusters are approximately half those of the bulk, and closer to the formation energies of substitutional defects. Despite the lower formation energies of S+I and 2S+I clusters on the $TiO_2$ surface, they remain higher than those of substitutional defects, explaining the tendency for small amounts of $SnO_x$ clusters to be formed. For the check of possibility of formation of $SnO_x$ clusters we performed the calculations for various distant substitutional (1S or 2S) and interstitial (I) impurities and find that formation energy increase in both case at the value of order 0.4 eV after move of interstitial atom in the next unit cell from 1S or 2S defects and further increase of distance also provide insignificant grow of formation energy. Therefore we can conclude that all types of Sn impurities on the surface of TiO2 demonstrate tendency to clusterisation.

**Table 1** Formation energies [eV/Sn atom] calculated for various types of bulk and surface defects (see Fig. 2).

| Type of defect | ZnO bulk | ZnO surface | TiO$_2$ bulk | TiO$_2$ surface |
|---|---|---|---|---|
| 1S | 0.74 | 0.63 | 2.13 | 1.02 |
| 2S | 1.02 | 1.32 | 2.42 | 0.43 |
| S+I | 3.04 | 2.51 | 2.57 | 1.27 |
| 2S+I | 2.72 | 2.65 | 2.76 | 1.87 |

We also have checked the effect of the most probable defects to the electronic structure and especially energy gap value. In the case of single substitutional impurity (1S) in the bulk or surface of ZnO occurs visible decreasing of the energy gap value (Fig. 7a). Note that employed method provides overestimation of energy gap value [28] and we can discuss only qualitative changes. In the case of TiO$_2$ host (Fig. 7b) 1S impurities in the bulk provides insignificant decreasing of energy gap value in contrast to the impurity incorporation in the surface when different configurations of substitutional impurities (1S and 2S) provide valuable decreasing of energy gap at almost the same value and formation of pair of substitutional and interstitial impurities (S+I) on the surface provides the appearance of the states on Fermi level similar to the previously discussed metallization by nS+I cluster formation in transitional metal doped TiO$_2$ [29] and ZnO [30].

**4. Conclusions**

ZnO:Sn and TiO$_2$:Sn matrices have been studied with XPS core-level and valence band spectroscopy. It was found that Sn is isovalently incorporated into zinc oxide (Sn$^{2+}$ → Zn$^{2+}$) and into titanium dioxide ($Sn^{4+} \rightarrow Ti^{4+}$), for both their bulk and film morphologies. But for $TiO_2$ films, the implantation leds to the clustering of Sn atoms, which did not occur in ZnO:Sn film samples. DFT calculations have shown that the formation energies of mixed S+I and 2S+I structural configurations on the surface of $TiO_2$ are higher than those of S and 2S substitutional defects, supporting our XPS observations of Sn metal clustering as well as the tendency to form some amount of $SnO_x$ clusters. In ZnO the opposite situation occurs: Sn dopants become incorporated into bulk and film ZnO hosts exclusively by means of substitution as single impurity centres.

**Acknowledgement**

The paper is supported by Ural Branch of Russian Academy of Sciences (Project 15-17-2-15), Russian Science Foundation for Basic Research (Project 13-08-00059) and Russian Federation Ministry of Science and Education (Government Task).